\documentstyle[pra,aps,preprint]{revtex}
\begin{document}
\draft
\title{Electron Correlation and Jahn-Teller Interaction in Manganese Oxides}
\author{Naoto Nagaosa, Shuichi Murakami}
\address{Department of Applied Physics, University of Tokyo,
Bunkyo-ku, Tokyo 113, Japan}
\author{Hyun Cheol Lee}
\address{Asia Pacific Center for Theoretical Physics, Seoul 151-742, Korea}
\date{\today}
\maketitle
\begin{abstract}
The interplay between the electron repulsion $U$ and the Jahn-Teller
electron-phonon interation $E_{\rm LR}$ is studied with a large $d$ model 
for the ferromagnetic state of the manganese oxides. 
These two interactions collaborate to induce the 
local isospin (orbital) moments and reduce the bandwidth $B$.
Especially the retardation effect of the Jahn-Teller phonon 
with the frequency 
$\Omega$ is effective to reduce $B$, but the strong $\Omega$-dependence
occurs even when the Coulombic interaction is dominating 
( $ U >> E_{\rm LR}$ ) as long as 
$E_{\rm LR} > \Omega$. The phonon spectrum consists of two components,
i.e., the temperature independent sharp peak at 
$\omega = {\tilde \Omega} = \Omega [(U +4 E_{\rm LR})/U]^{1/2}$
and that corresponding to the Kondo peak.
These results compared with the experiments suggest that 
$\Omega <E_{\rm LR} <U$ in the metallic manganese oxides.
\end{abstract}

\narrowtext
Since the discovery of colossal magnetoresistance
(CMR) the manganese oxides 
Re$_{1-x}$A$_x$MnO$_3$ ( Re: rare earth metal ion, A: divalent metal ion )  
have been attracting intensive interest recently \cite{exp}. 
One of the controversial issues is the role of orbital degeneracy 
of $e_g$-electrons. In the conventional models of these compounds
\cite{zener,anderson,degennes,furukawa},
the strong Hund's coupling is considered to be of  primary 
importance  and the orbital degeneracy has been often neglected. 
Recently, however,  Millis {\it et al.} \cite{millis,millis2} correctly 
pointed out that the Hund's coupling alone is 
not enough to explain the CMR together with the insulating temperature 
dependence of the resistivity $\rho(T)$ above the ferromagnetic transition 
temperature $T_c$. The additional coupling they proposed to be important is 
the Jahn-Teller electron-phonon coupling, which lifts the double degeneracy 
of $e_g$ orbitals and gives rise to the polaronic effect.
The small polaron formation above $T_c$ leads to the insulating 
$\rho(T)$. In contrast with the insulating phase, the spin alignment 
below $T_c$ will enlarge the band width $B$ as 
$B \propto \cos( \theta_{ij}/2)$ ($\theta_{ij}$  is the angle between the 
two neighboring spins 
${\vec S}_i$ and ${\vec S}_j$), and the Jahn-Teller interaction 
enters the weak coupling regime. This scenario seems to be 
consistent with the large isotope effect on $T_c$
by replacing O$^{16}$ by O$^{18}$ \cite{isotope}. 

The strong correlation models of  these compounds
have been also studied by several authors especially for the 
undoped cases \cite{kugel,ishihara}. 
In these models the Jahn-Teller interaction is considered to be 
smaller than the Coulomb interactions, and 
the effective Hamiltonian for the spin-orbital coupled system
and its phase diagram have been clarified. 
It is rather natural to assume the strong electron correlation because the 
strong Hund's coupling originates from
the strong electron-electron interactions ( $ \sim 5 {\rm eV}$ ). 
However, it is not trivial if the electron-electron interaction
continues to be strong even in the effective Hamiltonian describing 
the low energy physics after the screening by oxygen orbitals and 
conduction electrons, and it still remains the 
controversial issue whether the electron-electron interation and/or the 
Jahn-Teller coupling are in the strong coupling regime or not.

Experimentally there are several anomalous features 
which cannot be explained by the weak
Jahn-Teller coupling described above even in the low temperature 
ferromagnetic phase below $T_c$.

\noindent
[a] In the neutron scattering experiment 
no temperature dependent phonon modes have been observed   \cite{endoh}. 
The recent Raman scattering experiment also shows that 
the Jahn-Teller phonons ( especially their frequencies )
are temperature independent and 
insensitive to the ferromagnetic transition at $T_c$ \cite{yamamoto}.

\noindent
[b] The photo-emission spectra show a small discontinuity at
 the Fermi edge even at $T<<T_c$, which suggests some interactions 
 still remain strong there \cite{sarma}. 

\noindent
[c] The optical conductivity $\sigma(\omega)$ at $T<<T_c$ is composed of
two components, i.e., the narrow Drude peak ($\omega< 0.02 {\rm eV} $)
and the broad incoherent component extending up to $\omega \sim 1 {\rm eV}$
\cite{okimoto}.
The Drude weight is very small, which seems to be consistent with the 
photo-emission spectra.

\noindent
[d] The low temperature resistivity 
$\rho(T)$ can be fitted by
\begin{equation}
\rho(T) = \rho_0 + A T^2,
\end{equation}
where $A$ is a large constant of the 
order of 500$\mu \Omega {\rm cm}/ {\rm K}^2$ \cite{urusibara}, 
again suggesting the strong electron correlation.

\noindent
[e] Contrary to the case of resistivity [d],  the 
coefficient of $T$-linear specific heat is very small 
with $\gamma = 2{\rm mJ/K}$, which violates the Kadowaki-Woods law 
for these compounds \cite{hinetsu}.

Although [e] is difficult to reconcile with [a]-[d], we consider the latter
as the evidences for the strong coupling even at $T \ll T_c$.
Because the spins are perfectly aligned at $T \ll T_c$, 
the only remaining degrees of freedom 
  are the orbital ones. In this paper, we study  a
large $d$ model \cite{larged} for the ferromagnetic state 
including both the electron-electron interaction $U$ and the 
Jahn-Teller coupling $g$. This is the generalization of Ref.[7] 
in two respects, 
(i) including the electron-electron interation, (ii)
including the quantum fluctuations. Especially the latter is 
essential to describe the low temperature Fermi liquid 
state, which is described by the Kondo peak in the  large $d$ limit 
\cite{larged}.
The strong electron-electron interaction with the 
reasonable magnitude of the Jahn-Teller coupling explains both the 
large isotope effect and [a]. Moreover, the features of strong correlation 
[b]-[d] are at least consistent with the large $U$ picture although 
[e] still requires further studies,  which we have not undertaken in this paper.

 We start with the Hubbard-Holstein model for the ferromagnetic state. 
\begin{eqnarray}
H&=&-\sum_{i,j, \alpha, \alpha'} 
    t^{\alpha \alpha'}_{ij} c_{i \alpha}^\dagger  c_{j \alpha'}  
   +U \sum_i n_{i \uparrow} n_{i \downarrow} 
 \nonumber \\
  &-& g \sum_i Q_i ( n_{i \uparrow} - n_{i \downarrow} )
 + { 1 \over 2} \sum_i ( {{P_i^2} \over M}  + M \Omega^2 Q_i^2 ),
\end{eqnarray}
where the spin indices $\alpha = \uparrow, \downarrow$ correspond to the 
orbital degrees of freedom as $\uparrow = d_{x^2-y^2}$ and 
$\downarrow = d_{3z^2-r^2}$, and the real spins are assumed to be perfectly 
aligned. We consider only one  Jahn-Teller displacement mode $Q_i$ for each 
site, while there are two modes $Q_{2i},Q_{3i}$ in the real perovskite
structure \cite{kanamori}. 
However this does not change the qualitative features obtained
below. The Jahn-Teller mode is assumed to be an Einstein phonon with
a frequency $\Omega$. The transfer integral 
$t^{\alpha \alpha'}_{ij}$ depends on a pair of  orbitals 
$(\alpha,\alpha')$ and the hopping direction. These dependences lead to 
the various low lying orbital configurations, and 
thus they
 suppress  the orbital orderings in the ferromagnetic state.
Actually there are no experimental evidences for the orbital orderings,
e.g., the anisotropies of the lattice constants and/or transport properties 
in the ferromagnetic state. Then we assume that no orbital 
ordering occurs down to the zero temperature due to the quantum fluctuations,
and the transfer integral is assumed to be diagonal for simplicity, i.e.,
$t^{\alpha \alpha'}_{ij} = t_{ij} \delta_{\alpha \alpha'}$.
This means that the ground state is a Fermi liquid with two degenerate 
bands at the Fermi energy. In order to describe this Fermi liquid state,
we employ the large-$d$ approach where the model Eq.(2) is mapped to the 
impurity Anderson model with the self-consistent condition \cite{larged}. 
This approach has 
been applied to the manganese oxides to study the Hund's coupling by 
Furukawa \cite{furukawa} and to study the additional Jahn-Teller 
coupling  by Millis {\it et al.}  \cite{millis2}.
The action for the effective impurity model is given by
\begin{eqnarray}
S &=& \int^{\beta}_0  d \tau \int_0^\beta  d \tau' 
c^\dagger_{\alpha}(\tau)  G_0^{-1}(\tau-\tau')    c_{\alpha}(\tau') 
 \nonumber \\
&+&  \int_0^\beta  d \tau  \biggl[ U n_{\uparrow}(\tau) n_{\downarrow}(\tau) 
- g Q(\tau) ( n_{ \uparrow}(\tau) - n_{ \downarrow}(\tau) ) \biggr]
+ \int_0^\beta  d \tau  { M \over 2} 
\biggl[ (\partial_\tau Q(\tau) )^2 + \Omega^2 Q(\tau)^2 \biggr],
\end{eqnarray}
where 
$G_0^{-1}(\tau-\tau')$ is the dynamical Weiss field representing the 
influence from the surrounding sites. The self-consistency condition is 
that the on-site Green's function 
$G(i \omega_n) = ( G_0(i \omega_n)^{-1}
- \Sigma(i \omega_n) )^{-1}$  should be equal to the Hilbert 
transform of the density of states $D(\varepsilon)$ as
\begin{equation}
G(i \omega_n) = \int d \varepsilon \,
{ { D(\varepsilon)} \over 
{ i \omega_n + \mu - \Sigma(i \omega_n) - \varepsilon } }.
\end{equation}
We take the Lorentzian density of states  
$D(\varepsilon) = t/(\pi ( \varepsilon^2 + t^2))$ because 
there is no need to solve the self-consistency equation in this case, and
the Weiss field is given by 
\begin{equation}
G^{-1}_0(i \omega_n) =  i \omega_n + \mu + i\, t \,{\rm sign} \,\omega_n. 
\end{equation}
The only unknown quantity is the chemical potential $\mu$,
which is determined by the electron number. 
The determination of chemical potential
 requires some numerical calculations, but our discussion below is not
sensitive to the location of the chemical potential. Then we take $\mu$ as the 
parameter of our model, and the problem is now completely reduced  to that of 
a single impurity.
Now we introduce a  Stratonovich-Hubbard (SH) field $\xi(\tau)$ to 
represent the Coulomb interaction $U$. Then the 
action is given by
\begin{eqnarray}
\FL
& &S =\sum_{\omega_n}  
 (i \omega_n + \mu + i \,t \,{\rm sign} \,\omega_n )
c^\dagger_{\alpha} (\omega_n) c_{\alpha}(\omega_n) 
 \nonumber \\
& &+  \int_0^\beta  d \tau  \biggl[ 
{ {\Delta^2} \over U} \xi(\tau)^2 + 
{ M \over 2} [ (\partial_\tau Q(\tau) )^2 + \Omega^2 Q(\tau)^2 ]
+ \zeta(\tau) Q(\tau)
- ( \Delta \xi(\tau) + g  Q(\tau) ) 
( n_{ \uparrow}(\tau) - n_{ \downarrow}(\tau) ) \biggr],
\end{eqnarray}
where 
$\Delta = ( t^2 + \mu^2)/t$ and 
$\zeta(\tau)$ is the test field to measure the phonon correlation function.
At this stage the electron is coupled with the linear combination of the 
SH field $\xi$ and the phonon $Q$ as
$\eta(\tau) \equiv \xi(\tau) + {g \over \Delta}  Q(\tau)$. 
Then the effective action can be derived as
\begin{eqnarray}
S &=& \sum_{\omega_n} \biggl[ 
(i \omega_n + \mu + i t {\rm sign} \omega_n )
c^\dagger_{\alpha} (\omega_n) c_{\alpha}(\omega_n) 
- \Delta \eta(i \omega_n) 
( n_{ \uparrow}(- i \omega_n ) - n_{ \downarrow}(- i \omega_n) ) \biggr] 
\nonumber \\
&+&  \sum_{\omega_n} { 1 \over {2M(\omega_n^2 + {\tilde \Omega}^2 ) }}
\biggl[ 
- \zeta(i \omega_n) \zeta(- i \omega_n)
+ {{2 \Delta^2} \over U} M(\omega_n^2+ \Omega^2) 
\eta(i \omega_n) \eta(- i \omega_n)
\nonumber \\
&+& { {2 g \Delta} \over U }
( \zeta(i \omega_n) \eta(- i \omega_n)
+\eta(i \omega_n) \zeta(- i \omega_n) ) \biggr],
\end{eqnarray}
where ${\tilde \Omega} = \Omega \sqrt{U_{\rm eff.}/U}$.
The effective interaction $U_{\rm eff.}$ is the sum of the 
Coulomb repulsion $U$ and the lattice relaxation energy $E_{\rm LR} = 
g^2/(2M\Omega^2)$ as $U_{\rm eff.} = U + 4E_{\rm LR}$.
The phonon Green's function 
$d(i \omega_n) =  \langle Q(\ i \omega_n) Q(-i \omega_n) \rangle$
is given by 
\begin{equation}
d(i \omega_n) =  
{ 1 \over {M(\omega_n^2 + {\tilde \Omega}^2 ) }}
+ 4 \biggl( { {g \Delta / U}  \over { M(\omega_n^2 + {\tilde \Omega}^2 ) }}
\biggr)^2 \chi_\eta( i \omega_n),
\end{equation}
where 
$\chi_\eta( i \omega_n) = \langle \eta(i \omega_n) \eta(- i \omega_n) \rangle$ 
is the orbital susceptibility of the electronic system.
The first term is the usual phonon Green's  function
with the renormalized frequency ${\tilde \Omega}$, which is higher 
than the bare $\Omega$. This 
${\tilde \Omega}$ does not depend on the electron response function
and hence temperature independent. 
When $U \ll E_{\rm LR}$, ${\tilde \Omega}$ is much larger than $\Omega$
and the first term  of Eq.(8) is irrelevant in the phonon frequency region 
$\omega \sim \Omega \sim 0.1 {\rm eV}$.
In the opposite limit $U \gg  E_{\rm LR}$, the 
renormalization of the phonon frequency is small. 
The second term depends on the 
temperature (Kohn anomaly) and shows the characteristic lineshape with the
broadening.  
From the above  consideration together with the experimental fact that 
no temperture dependence of the phonon frequency has been observed, 
we conclude that the Coulomb interaction $U$ is larger than the lattice 
relaxation energy $E_{\rm LR}$.

Now let us study the electronic system in detail.
Following Hamann \cite{hamann}, 
we can obtain the solution of the integral equation for the
electron Green's function in the presence of the time-dependent 
field $\eta(\tau)$, and hence for the effective action.
The result is 
\begin{eqnarray}
S_{\rm eff.}   &=& 
 \sum_{\omega_n} 
{{\Delta^2} \over U} 
{ {\omega_n^2+ \Omega^2} 
 \over {\omega_n^2 + {\tilde \Omega}^2  }}
\eta(i \omega_n) \eta(- i \omega_n)
\nonumber \\
&-& \int_0^\beta  d \tau
 {{ 2 \Delta} \over \pi} 
\biggl[ \eta(\tau) \tan^{-1} \eta(\tau) - 
{ 1 \over 2} \ln ( 1 + \eta(\tau)^2)
\biggr]
\nonumber \\
&+& { 1 \over {\pi^2}} 
\int_0^\beta  d \tau
\int_0^\beta  d \tau'
P { 1 \over { \tau - \tau'}} \eta(\tau) 
{ { d \eta(\tau')} \over { d \tau'}}
{ 1 \over { \eta(\tau)^2 - \eta(\tau')^2}} 
\ln {{ 1 + \eta(\tau)^2} \over { 1 + \eta(\tau')^2} },
\end{eqnarray}
where $P$ denotes the principal value.
The Eq.(9) differs from Eqs.(3.40)-(3.43) of Ref.\cite{hamann} by the frequency
dependent coefficient of 
$ \eta(i \omega_n) \eta(- i \omega_n)$ due to the retardation effect of the 
phonons. A similar model without the Coulomb repulsion $U$ has been studied 
by Yu-Anderson \cite{yu}, which corresponds to the limit $U \to 0, 
{\tilde \Omega} \to \infty,   
{\tilde \Omega}^2 U = { \rm finite}$ in Eq.(9). 
First, we consider the saddle point solution 
for the $\eta$ field assuming the static configuration $\eta(\tau) = \eta_0$.
Then the action for the static solution becomes 
\begin{equation}
S_0 = \beta \biggl[ 
{{ \Delta^2} \over {U_{\rm eff.}} } \eta_0^2
- {{2  \Delta} \over \pi} 
\biggl( \eta_0 \tan^{-1} \eta_0 - 
{ 1 \over 2} \ln ( 1 + \eta_0^2) \biggr) \biggr].
\end{equation}
When $U_{\rm eff.}/ ( \pi \Delta) <1$, there is a single minimum at the origin,
 and the double minima appear for 
 $U_{\rm eff.}/ ( \pi \Delta) >1$.
We are interested in the limit of strong correlation
 $U_{\rm eff.}/ ( \pi \Delta) >>1$, and in this limit the tunneling events
( instantons) between the two minima at $\eta =  \pm \eta_0
 \cong \pm { {U_{\rm eff.}} \over {2 \Delta}}$ play an essential role
in producing the Kondo peak at the Fermi level. 
The third term in Eq.(9) gives the 
non-local interaction between the instantons along the time axis, 
which corresponds to 
$J_z$ term when it is mapped to the Kondo problem.
The fugacity $z$ of the instanton, which corresponds to  $J_{\perp}$,
can be estimated following ref.\cite{hamann}. 
Let $\tau_0$ be the width of the instanton,
i.e., hopping time, and the action for a single instanton is given by 
\begin{equation}
S_{\rm inst.}(\tau_0)  = \sum_{\omega_n}  
{ { 2 g^2 \Delta^2 \omega_n^2}  \over 
{U^2 M (\omega_n^2 + {\tilde \Omega}^2 ) 
{\tilde \Omega}^2  }}\,\,
\eta_{\rm inst.}(i \omega_n) \eta_{\rm inst.}(- i \omega_n)
+ { 1 \over 6} \tau_0 U_{\rm eff.} - \ln \tau_0.
\end{equation}
The last term can be written in terms of the derivative 
$h(\tau) = d \eta_{\rm inst.}(\tau)/ d \tau$, which is the 
localized function around the center of the instanton ( $\tau=0$)
with the width of the order of $\tau_0$ and the integral of $ h(\tau)$
over $\tau$ is 
$2 \,\eta_0$. We assume 
$h(\tau) = ( \eta_0/\tau_0) e^{- |\tau|/\tau_0}$. Then 
the summation over $\omega_n$ can be easily done and we obtain
\begin{equation}
S_{\rm inst.}(\tau_0)  = 
{ {4 E_{\rm LR}} \over {\Omega^2 \tau_0 } } f({\tilde \Omega} \tau_0)
+ { 1 \over 6} \tau_0 U_{\rm eff.} - \ln \tau_0,
\end{equation}
where $f(\alpha) = \alpha (\alpha + 2)/(4(1+ \alpha )^2)$. Minimizing 
$S_{\rm inst.}(\tau_0)$ with respect to $\tau_0$, we obtain $\tau_0$ as 
$\tau_0 \cong 6/U_{\rm eff.}$ for $U_{\rm eff.} >> 6 {\tilde \Omega}$
and 
$\tau_0 \cong 3/U_{\rm eff.}
+ \sqrt{ (3/U_{\rm eff.})^2 + 6 E_{\rm LR}/(\Omega^2 U_{\rm eff.}) }$
for $U_{\rm eff.} << 6 {\tilde \Omega}$. Introducing the dimensionless
variables  $x = E_{\rm LR}/\Omega$ and $y = U/E_{\rm LR}$, 
the reduction factor $R$ for the instanton
fugacity $z =  \eta_0^{-1} e^{-R} $ is given by
\begin{eqnarray}
R &=& 2x \sqrt{ 1 + 4 y^{-1}},\quad {\rm for}\;\; y >> 1/\max(x,x^2) \\
R& =& \sqrt{ 1 + { 2 \over 3} x^2(4+y) }, \quad {\rm for} \;\; y << 1/\max(x,x^2).
\end{eqnarray}
The main conclusion here is that when $x>1$, i.e., $E_{\rm LR} > \Omega$,
the reduction factor $R$ is proportional to   $x$ even in the limit
 $U \gg E_{\rm LR}$.
This is because the overlap of the phonon wavefunction enters the 
tunneling matrix element even when the Coulomb interaction is dominating.

In summary, the fugacity of the instantons and the 
long range interaction between them can be mapped to the 
anisotropic Kondo model with the
Kondo couplings given by 
\begin{eqnarray}
N(0)J_{\perp} &\cong& { {\Delta} \over { U_{\rm eff.} } }e^{-R}
\nonumber \\
N(0)J_z &\cong& { {\Delta} \over { U_{\rm eff.} } },
\end{eqnarray}
where $N(0)$ is the density of states for the Kondo problem and the
reduction factor $R$ is estimated to be Eqs.(13) and (14).
Solving the scaling equations for $J_{\perp}$ and $J_z$, an estimate of the
Kondo temperature $T_K$ is obtained for large $R (>>1)$ as
\begin{equation}
T_K \sim t  \exp\biggl[ - {{RU_{\rm eff.}} \over { \Delta}} \biggr].
\end{equation}
This Kondo temperature gives the effective bandwidth $B$ in
the large-$d$ model, and the effective mass enhancement is 
estimated as $m^*/m = t/T_K \sim
\exp\biggl[ {{RU_{\rm eff.}} \over { \Delta}} \biggr]$.
This strong mass enhancement manifests itself in the physical quantities as follows.
The orbital susceptibility $\chi_{\eta}(i \omega_n \to \omega + i \delta)$
has a peak at $\omega=0$ with the peak height of the order of 
$\eta_0^2/T_K$ and with the width of the order of $T_K$.
The specific heat  coefficient $\gamma$ should be proportional to 
$B^{-1}$, while the coefficient $A$ of $T^2$ in Eq.(1) scales as
$A \propto B^{-2}$. 
The ferromagnetic transition temperature $T_c$ is expected 
to be proportional to $B$, and hence we expect a large  
isotope effect when $R>1$.
In other words, the large isotope effect does not rule out the large 
Coulomb interaction $U$.
The edge of the photoemission spectrum at the Fermi energy and the 
Drude weight are reduced by the factor $m/m^*$.
As mentioned above, many experiments except for those of  
specific heat seem to suggest that the manganese oxides in their
ferromagnetic state belong to the strongly correlated region.
Then it is natural that the large isotope effect is observed.
The band width is rather difficult to pin down accurately, but a
rough estimate from $A$ in eq.(1) is 
$B \sim 0.1 {\rm eV}\,$, which is comparable to $\Omega$.
This suggests that the suppression is not extremely large. Another 
experimental fact is that the temperature independent Jahn-Teller 
phonon is observed at the frequency near the original one \cite{yamamoto}, 
which means that $U$ is larger or at least of the order of $E_{\rm LR}$.
Then we conclude that the relative magnitude of the 
energy scales is $\Omega<E_{\rm LR}<U$ for the ferromagnetic phase in the 
manganese oxides.
The small specific heat coefficient necessitates 
the physical mechanisms which have not been included 
in our model.  One possibility is the inter-site process like the 
orbital singlet formation, and another possibility is that 
the proximate quantum criticality  is affecting the Fermi liquid 
behavior. 

In summary, we have studied a model of manganese oxides in the ferromagnetic 
state taking into account both the Coulomb repulsion $U$ and the Jahn-Teller 
interaction $E_{\rm LR}$. These two interactions collaborate to induce the 
local orbital moments.
In the strong coupling case, i.e., $U_{\rm eff.} =U+4E_{\rm LR}>> t$
($t:$ transfer integral), the overlap of the phonon wavefunctions
($\sim e^{-E_{\rm LR}/\Omega}$) enters the tunneling amplitude between the 
two minima for the orbital moment.  
The phonon spectral function consists of two parts, i.e., 
the sharp peak at the renormalized frequency ${\tilde \Omega}
= \Omega \sqrt{ U_{\rm eff.}/U}$ and the broad peak with the 
width of the order of the band width.
These results can reconcile the
large isotope effect on $T_c$ and  the apparent temperature independent 
phonon spectrum assuming $U > E_{\rm LR}>\Omega$.

The authours would like to thank  
Y.Tokura, A. Millis, J. Ye, G. Kotliar, Q. Si, 
H. Kuwahara, S. Maekawa, H.Fukuyama, K. Miyake, S. Ishihara, 
T.Okuda, K.Yamamoto and Y.Endoh for valuable discussions. 
This work was supported by COE and Priority Areas Grants from 
the Ministry of Education, Science and Culture of Japan.
Part of this work has been done in APCTP workshop and
Aspen Center for Physics, and we acknowledge their
hospitalities.

\end{document}